\documentclass[12pt, draftclsnofoot, onecolumn]{IEEEtran}
\usepackage[latin1]{inputenc}
\usepackage{booktabs}
\usepackage{multirow}
\usepackage{color}
\usepackage{mathrsfs}
\usepackage{cite}
\usepackage{amsmath,amssymb,amsthm}
\usepackage{mathtools}
\usepackage{subfigure}
\newtheorem{lemma}{Lemma}
\usepackage{bm}
\newtheorem{theorem}{Theorem}

\usepackage{epsfig}
\usepackage{dblfloatfix}

\DeclareMathOperator*{\argmax}{\arg\max}
\DeclareMathOperator{\Hessian}{Hess}
\DeclareMathOperator\erf{erf}

\usepackage{color,soul}
\definecolor{laranja}{rgb}{1.0, 0.6, 0.2}

\begin{document}

\title{Maximum Secrecy Throughput of MIMOME FSO Communications with Outage Constraints}
\author{Marcos~Eduardo~Pivaro~Monteiro,~\IEEEmembership{Student~Member,~IEEE,} João~Luiz~Rebelatto,~\IEEEmembership{Member,~IEEE,} Richard~Demo~Souza,~\IEEEmembership{Senior~Member,~IEEE,}
and~Glauber~Brante,~\IEEEmembership{Member,~IEEE}
\thanks{Copyright (c) 2017 IEEE. Personal use of this material is permitted. However, permission to use this material for any other purposes must be obtained from the IEEE by sending a request to pubs-permissions@ieee.org.}
\thanks{This work has been partially supported by CNPq and CAPES (Brazil).}
\thanks{Marcos Eduardo Pivaro Monteiro, João Luiz Rebelatto and Glauber Brante are with CPGEI, UTFPR, Curitiba, PR, 80230-901, Brazil (e-mail: marcosmonteiro@alunos.utfpr.edu.br, \{jlrebelatto, gbrante\}@utfpr.edu.br).}
\thanks{Richard Demo Souza is with Federal University of Santa Catarina (UFSC), Florianópolis, SC, 88040-900, Brazil (e-mail: richard.demo@ufsc.br).}
}
%\author{Marcos~Eduardo~Pivaro~Monteiro,\textsuperscript{1}~\IEEEmembership{Student~Member,~IEEE,}\\ João~Luiz~Rebelatto,\textsuperscript{1}~\IEEEmembership{Member,~IEEE,}\\Richard~Demo~Souza,\textsuperscript{2}~\IEEEmembership{Senior~Member,~IEEE,}\\
%and~Glauber~Brante,\textsuperscript{1}~\IEEEmembership{Member,~IEEE}
%\thanks{Copyright (c) 2017 IEEE. Personal use of this material is permitted. However, permission to use this material for any other purposes must be obtained from the IEEE by sending a request to pubs-permissions@ieee.org.}
%\thanks{This work has been partially supported by CNPq and CAPES (Brazil).}
%\thanks{\textsuperscript{1}Graduate Program in Electrical and Computer Engineering, Federal University of Technology, Curitiba, PR, $80230-901$, Brazil\*
%\textsuperscript{2}Federal University of Santa Catarina, Florianópolis, SC, $88040-900$, Brazil}
%}
%\affil{\textsuperscript{1}Graduate Program in Electrical and Computer Engineering, Federal University of Technology, Curitiba, PR, 80230-901, Brazil\\\textsuperscript{2}Federal University of Santa Catarina, Florianópolis, SC, 88040-900, Brazil}

\maketitle

%\markboth{IEEE Photonics Journal}{ }

\begin{abstract}
In this work, we consider a scenario where two multiple-aperture legitimate nodes (Alice and Bob) communicate by means of Free-Space Optical (FSO) communication in the presence of a multiple-aperture eavesdropper (Eve), which is subject to pointing errors. Two different schemes are considered depending on the availability of channel state information (CSI) at Alice: {\it i)} the adaptive scheme, where Alice possesses the instantaneous CSI with respect to Bob;  {\it ii)} the fixed-rate scheme, where such information is not available at Alice. The performance of the aforementioned schemes is evaluated in terms of a recently proposed metric named effective secrecy throughput (EST), which encompasses both the reliability and secrecy constraints. By constraining the system to operate below a given maximum allowed secrecy outage probability, we evaluate the EST analytically and through numerical results, showing that the use of multiple apertures at Alice is very important towards achieving the optimal EST. %Furthermore, we demonstrate that, by exploiting the pointing errors in the eavesdropper channel, it is possible to significantly increase the EST.
%In this paper, we evaluate the EST of a communication between two legitimate peers (Alice and Bob) in a Multiple-Input-Multiple-Output (MIMO) point-to-point Free-Space Optical (FSO) communication in the presence of an eavesdropper (Eve). Following a recently proposed work, we consider that the system cannot operate above a given maximum allowed secrecy outage probability (SOP). We analyzed the performance in terms of EST with eavesdropper outage constraints for two schemes, namely adaptive and fixed-rate. For the adaptive scheme, the instantaneous capacity of the legitimate channel is available at the transmitter, while for the fixed-rate scheme, the legitimate channel instantaneous capacity is not available. Through the use of analytical and numerical results, we demonstrate the importance of the use of multiple apertures in order to maximize the EST and minimize the leaked information. We also obtain closed form expressions for both the SOP and the EST for both schemes, which are useful tools to a system designer avoiding time consuming simulations.
\end{abstract}


\section{Introduction} \label{sec:introducao}
\bstctlcite{biblio:BSTcontrol}

Secrecy in the presence of an eavesdropper is a classical problem in communication theory, which was first analyzed in the context of the wiretap channel in~\cite{Wyner1975}, and received renewed interest in radio-frequency (RF) wireless applications in the past few years due to their broadcast nature. Recently, physical layer security has appeared as a complement to the cryptographic techniques~\cite{barros.06.secrecy}, showing that the fading, usually a negative factor in terms of reliability, can be used to increase the data security. In the case of free-space optical (FSO) transmissions, security is intrinsically higher than in RF scenarios due to the high directionality of optical beams. However, the interception of optical signals is also possible and efforts must be expended aiming to avoid it.

In order to intercept the legitimate link, the eavesdropper (Eve) may either approach the legitimate transmitter (Alice) and try to block the laser beam in order to collect a large amount of power, or approach the legitimate receiver (Bob) and take advantage of the beam radiation being reflected by small particles, receiving part of the signal intended for Bob. The second case is more reasonable as a real threat scenario since, if Eve is close to Alice, she will not be able to intercept the beam without blocking the line of sight, which could allow Alice to detect its presence visually or based on the variation of the received power experienced by Bob~\cite{Martinez2015}.

The use of wiretap codes~\cite{bloch.11} is the usual assumption to achieve secrecy capacity, for which a redundancy rate is defined as $\mathcal{R}_E = \mathcal{R}_B - \mathcal{R}$, where $\mathcal{R}$ represents the target secrecy rate and $\mathcal{R}_B$ corresponds to the rate of transmitted codewords. Then, reliable and secure communication requires that: {\it i)} $\mathcal{R}_B \leq C_B$ (reliability constraint), where $C_B$ is the instantaneous channel capacity of the legitimate link; {\it ii)} $\mathcal{R}_E > C_E$ (secrecy constraint), where $C_E$ is the instantaneous channel capacity between Alice and Eve. However, it is very unlikely that Alice has knowledge about the instantaneous channel state information (CSI) with respect to Eve~\cite{khisti.10} and, thus, the condition $\mathcal{R}_E > C_E$ cannot be guaranteed at all times. In this particular case, one must resort a probabilistic analysis through the secrecy outage probability (SOP)~\cite{barros.06.secrecy}.

For instance, the effective secrecy throughput (EST) was proposed in~\cite{yan.2015} as a way to encompass both reliability and secrecy in a single metric. The EST is adopted in order to evaluate the performance of a RF-based single-input single-output multiple-antenna eavesdropper scenario considering the so-called \emph{adaptive} and \emph{fixed-rate} schemes. For the adaptive scheme, $C_B$ is required to allow Alice to adapt $\mathcal{R}_E$ and $\mathcal{R}$ accordingly, guaranteeing the reliability constraint. On the other hand, only the expected value of $C_B$ is assumed at Alice for the fixed-rate scheme. Moreover, the optimal secrecy rate that maximizes the EST is also investigated in~\cite{yan.2015}. However, such optimization is not constrained with respect to the SOP, which means that the optimal EST provided by~\cite{yan.2015} may lead to an outage probability that might be above an acceptable security threshold. Aiming at avoiding this possible security issue, the EST was extended in~\cite{Monteiro2015} with the addition of a constraint on the maximum allowed SOP. Such constrained EST was then adopted to evaluate the performance of a RF-based multiple-input multiple-output (MIMO) multiple-antenna eavesdropper system, subjected to Rayleigh fading and operating under the adaptive scheme from~\cite{yan.2015}.

It is also worthy mentioning that the secrecy can be improved by using techniques such as orbital angular momentum (OAM) multiplexing, scintillation reciprocity and acousto-optic deflectors. The use of orbital angular momentum multiplexing (OAM) is studied in~\cite{Sun.2016} to increase the aggregate secrecy capacity\footnote{The aggregate secrecy capacity is defined as the summation of the secrecy capacity for all multiplexed channels~\cite{Sun.2016}.}, and it is demonstrated that the performance can be improved for weak and medium turbulence regimes. In~\cite{Eghbal.2014}, the use of acousto-optic deflectors are proposed to further increase the data security. In such approach, optical messages are sent through different beam paths between Alice and Bob, while it is shown that the radius of the beam, and the intensity of the received beam from different beam paths, directly affects the data transmission security. Moreover, an air-to-ground FSO communication system is investigated in~\cite{Wang.2014}, demonstrating that, for any FSO system where the scintillation reciprocity holds, the communication can be further improved through the use of a cryptosystem relying on the securely generated keys. While each of the aforementioned methods can be used to further improve the FSO communication in its own way, none of them are based on wiretap codes (i.e., do not use the redundancy rate) and, thus,  does not give any insight about the optimum value of $\mathcal{R}_E$.

In this work, we consider a multiple-input multiple-output multi-apertures eavesdropper (MIMOME) coherent FSO scenario. We assume that Alice adopts a transmit laser selection (TLS) scheme, which was shown to improve reliability when adding more apertures between the transmitter and the receiver~\cite{Rjeily2015}, while Bob and Eve operate under the optimal maximum ratio combining (MRC) scheme~\cite{goldsmith.05}. By adopting the constrained EST metric from~\cite{Monteiro2015} and considering gamma-gamma distribution to model the fading of the FSO channel, we evaluate the performance of the adaptive and fixed-rate transmission schemes from~\cite{yan.2015}. Furthermore, we also assume that the legitimate link between Alice and Bob experiences no misalignment issues, which generally holds in scenarios when the receiving apertures are located fairly close to each other~\cite{Sandalidis.2009}. As a consequence, the fraction of the power received by Eve also encompasses the existence of pointing errors, which generally depends on the distance between Bob and Eve~\cite{Sandalidis.2009,Bhatnagar.16,Farid.2007}. The main contributions of this paper are described as follows:
\begin{enumerate}
	\item We obtain closed form EST expressions for both adaptive and fixed-rate schemes in a coherent FSO scenario where the eavesdropper is subject to pointing errors, which are verified by numerical results;
	\item The rates $\mathcal{R}_E$ and $\mathcal{R}_B$ that maximize the EST for both adaptive and fixed-rate schemes are analytically obtained, respecting the constraint of a maximum allowed SOP;
	\item We demonstrate that, when operating under the fixed-rate scheme, including additional apertures can lead to a larger EST than that obtained using the adaptive scheme.
%	\item The pointing errors, which represents the misalignment in the FSO link, are used to model the eavesdropper channel. It is shown that such realistic approach improves the EST and should be taken into account in order to obtain the optimum values of $\mathcal{R}_E$ and $\mathcal{R}_B$.
\end{enumerate}
%(between Alice and Bob better)

The rest of this paper is structured as follows. Section~\ref{sec:preliminaries} presents the system model, the adaptive and fixed-rate transmission schemes, and the EST performance metric. Considering a MIMOME FSO communication,  Sections~\ref{sec:mimome_fso_adap} and~\ref{sec:mimome_fso_fixe} present the performance analysis of the adaptive and the fixed-rate transmission schemes, respectively. Section~\ref{sec:numerical_results} presents numerical results, while Section~\ref{sec:conclusions} concludes the paper.
%%#############################################################################
\section{Preliminaries} \label{sec:preliminaries}

\subsection{System Model}

\begin{figure} [!t]
\begin{center}
\includegraphics[width=240px]{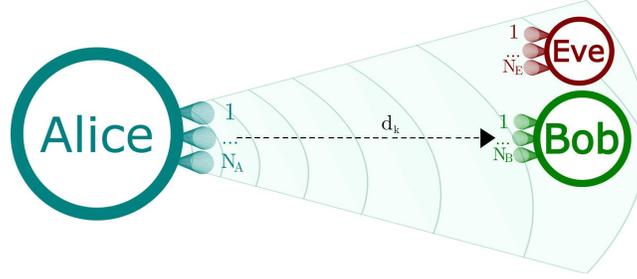}%width=0.35\textwidth
\end{center}
\caption{MIMOME point-to-point FSO communication, composed of a legitimate transmitter (Alice, provided with $N_A$ transmit apertures) and a legitimate receiver (Bob, provided with $N_B$ receive apertures), communicating in the presence of an eavesdropper (Eve, with $N_E$ apertures).}
\label{fig:figMS}
\end{figure}

The model adopted in this work is composed of one legitimate transmitter, Alice ($A$), communicating with a legitimate receiver, Bob ($B$), in the presence of an eavesdropper, Eve ($E$). Alice is equipped with $N_A$ transmit apertures working under transmit laser selection (TLS) scheme, while Bob and Eve are provided with, respectively, $N_B$ and $N_E$ receive apertures, using the optimum MRC scheme. This scenario, referred to as MIMOME, is illustrated in Fig.~\ref{fig:figMS}.
%\footnote{Q-ary PPM is an interesting alternative to on-off keying (OOK), due to being more energy efficient for $Q > 2$~\cite{Xu.2009} while having the same energy efficient for $Q=2$, and IM/DD are used in most current FSO systems due to implementation complexity issues~\cite{Xu.2009}.}
%Following~\cite{Rjeily2015}, the FSO transmission considered employs Binary Pulse Position Modulation (BPPM) with Intensity-Modulation and Direct-Detection (IM/DD)\footnote{For high bandwidth scenarios, as the bit duration $T_b \rightarrow 0$, binary level techniques are capacity achieving~\cite{Hranilovic.2009,Khalighi.2014}.}.

%While direct detection-based FSO uses a photodetector for directly converting the optical signal power to alternating current (AC), in coherent FSO systems the received optical signal is combined with a local oscillator signal to produce an AC photocurrent  proportional to the optical signal power~\cite{Niu.2013}.
Depending on the detection type, FSO communication systems can be separated in two main categories, namely coherent and direct detection (DD) systems~\cite{Aghajanzadeh.2010}.  While the maximum capacity bounds of DD systems have been studied in a number of works such as~\cite{Chaaban.2016, Aghajanzadeh.2015}, in~\cite{Kumar.2015}, it is shown that coherent detection outperforms direct detection at the cost of a higher complexity. Furthermore, the extraction of phase information for coherent FSO systems allows a greater variety of modulation formats in comparison with direct detection~\cite{Niu.2013}.
%For direct detection, common modulation formats are on off keying (OOK) and pulse position modulation (PPM), while, for coherent detection, common modulation formats are phase shift keying (PSK), frequency shift keying (FSK), amplitude shift keying (ASK), quadrature shift keying (QPSK) and differential phase shift keying (DPSK)~\cite{Attri.2015}. Note that, even though the coherent detection offers superior performance, direct detection systems are low cost and simpler to implement.

%Differently from RF communication where the wireless channel can be modeled by means of easy-to-manipulate distributions such as Rayleigh or Nakagami-$m$~\cite{goldsmith.05}, 
The transmission in FSO communication is affected by a large number of phenomena, among which the most harmful is the scintillation, defined as the random fading characteristic of the received optical intensity~\cite{Martinez2015}. As a consequence, the received optical irradiance in FSO communication, regardless the detection type, is commonly modeled by means of more complex (and difficult to manipulate) statistical distributions such as gamma-gamma~\cite{Park.2015,Kahn1997,Habash2001}. Apart from scintillation effect, pointing errors must also be taken into account when there is a non-negligible misalignment between the transmitter and receiver nodes~\cite{Bhatnagar.16}. While in the legitimate link perfect alignment is commonly assumed~\cite{Sandalidis.2009}, such assumption is not realistic to the link between Alice and Eve, since Eve cannot be too close to Bob in order not to be detected.

Following~\cite{Fried.1967,Belmonte.2008,Aghajanzadeh.2010}, we employ coherent detection, which, despite being more complex than  direct detection, provides flexibility since either amplitude, frequency or phase can be used. In such systems, even though the capacity initially increases with the increase in the diameter of the receiver aperture, it tends to saturate, justifying the use of multiple apertures at the receiver~\cite{Kaur.2014}. We also consider that the irradiance received at apertures in Bob and Eve are independent, i.e., the large-scale and small-scale effects experienced by Bob are independent from that seen at Eve, which holds in a scenario where the distance between Bob and Eve are greater than the correlation length $d_0~\approx~\sqrt{\lambda d_k}$~\cite{Navidpour.2007}, where $\lambda$ is the wave length and $d_{k}$ is the distance between the transmit and receive $k~\in~\{B,E\}$ nodes~\cite{Rjeily2015}. In FSO communications through the turbulent atmosphere, the maximum achievable rate per unit of bandwidth is given by $\log_2(1 + \gamma_{k})$ bits/s/Hz, where $\gamma_{k}$ is the signal to noise ratio (SNR) at the receiver, which is random due to the nature of the channel~\cite{Belmonte.09}. If the noise is dominated by local oscillator shot noise, the SNR at Bob or Eve, after a transmission from Alice, can be expressed as~\cite{Niu.2012,Niu.2013,Niu.2009}%\footnote{Note that, thus, we assume that the received optical irradiances for both Bob and Eve are independently and identically distributed.}
\begin{equation}
\gamma_{k} = \frac{A_0 \eta_e E_s A}{h f_o \Delta_f} \mathcal{I}_{k},
\label{eq1}
\end{equation}where $\eta_e$ is the quantum efficiency of the photodetector, $E_s$ is the symbol energy, $A$ is the beam waist area, $h$ denotes the Planck's constant, $f_o$ denotes the frequency of the received optical signal, $\Delta_f$ is the is the noise equivalent bandwidth, $A_0~=~\erf^2(\nu )$  represents the fraction of the available power received at node $k$ for the photodetector area when there is no misalignment between the transmitter and the receiver, $\erf(\cdot)$ is the error function, $\nu = \sqrt{\pi {2}} \rho  / {\omega _b}$, $\rho$ is the radius of the receive aperture and $\omega _b$ is the received beam size.

Finally, $\mathcal{I}_k$ represents the irradiance associated with the link between Alice and receiver $k$. For a given $i$-th transmit aperture of Alice and $j$-th receive aperture of node $k$, $\mathcal{I}_k$ can be expressed as $\mathcal{I}_{k}^{i,j}=I_{l,k}^{i,j}~I_{a,k}^{i,j}~I_{p,k}^{i,j}$, where $I_{a,k}^{i,j}$ is the fading caused by atmospheric turbulence, $I_{p,k}^{i,j}$ is the pointing error and $I_{l,k}^{i,j}$ represents the attenuation due to path-loss. Similarly to RF wireless channels~\cite{Zhu.2010}, since large-scale fluctuations in the irradiance are generated due to turbulent eddies, we follow~\cite{Balsells.14} and assume that, for a given receiver node $k$, the large-scale effects are fully correlated among the receiving apertures, which holds when the received signal in each photodetector propagates through the same large-scale eddies\footnote{The assumptions of perfect alignment between Alice and Bob and the fully correlated large-scale fluctuations require that the number of transmitting and receiving apertures is not large, which is justified, respectively, by the space necessary to place each aperture and by the fact that large-scale fluctuations are produced by turbulent eddies with limited sizes, ranging from the scattering disk to the outer scale~\cite{Balsells.14}.}. Thus, without loss of generality, in the rest of this paper we assume that $I_{l,k}^{i,j}~=~1$.

The pointing errors in the legitimate link are assumed to be negligible, such that $I_{p,B}^{i,j}~=~1$. This can be achieved in practice, for example, by means of perfect alignment~\cite{Sandalidis.2009}. However, the same does not hold to Eve, which is subjected to pointing errors. We also consider that the receiving apertures of Eve are close enough such that all of them are affected in the occurrence of a pointing error.

Thus, for the MIMOME model adopted in this work, $\mathcal{I}_k$ can be written as~\cite{Niu.2013,Niu.2012,Zambrana2009,Rjeily2011}
\begin{equation}
\begin{split}
\mathcal{I}_{k} = \! \left\{ \!\!\!\!
\begin{array}{ll}
  \displaystyle I_{p,E} \sum^{N_{E}}_{j=1} I_{a,E}^{i,j}, &\qquad \!\!\!\text{Eve}; \\
 \displaystyle\max_{i=1,...,N_A} \displaystyle\sum^{N_{B}}_{j=1} I_{a,B}^{i,j}. &\qquad \!\!\!\text{Bob-TLS}. \\
\end{array}\right.
\label{eqtlsegc}
\end{split}
\end{equation}

Note that, although the eavesdropper is capable of accessing the feedback channel from Bob to Alice, the selected aperture is optimum to the legitimate channel only so that the aperture index alone cannot be exploited by the eavesdropper~\cite{alves.12.secrecy.tas}. This behavior is represented by the $\max_{i=1,...,N_A}$ term in \eqref{eqtlsegc}. Following~\cite{Tsiftsis.2008,Rjeily2015,Khalighi2009,Habash2001}, we also adopt the gamma-gamma fading model to represent the turbulence induced by scintillation, in which the pdf of the turbulence in a single link ($I_{a,k}^{i,j}$, which we refer to as $I$ in order to ease the notation) for $(I \geq 0)$ is given by
\begin{equation}\label{eq:pdfgammagamma}
f^{\gamma\gamma}(I) = \frac{2(\alpha\beta)^{(\alpha+\beta)/2}}{\Gamma(\alpha)\Gamma(\beta)} I^{(\alpha+\beta)/2-1}K_{\alpha-\beta}(2\sqrt{\alpha\beta I}),
\end{equation}
where $K_{c}(\cdot)$ is the modified Bessel function of the second kind and order $c$, and $\Gamma(\cdot)$ is the gamma function.

From~\cite{Rjeily2015}, the parameters $\alpha$ and $\beta$ are given by
\begin{subequations}
\begin{equation}
\label{eq:alpha}
\alpha(d_{k}) = \left[\exp\left( \frac{0.49\sigma^2_R (d_{k})}{(1 + 1.11 \sigma^{12/5}_R (d_{k}))^{7/6}} \right) - 1\right]^{-1},
\end{equation}
\begin{equation}
\label{eq:beta}
\beta(d_{k}) = \left[\exp\left( \frac{0.51\sigma^2_R (d_{k})}{(1 + 0.69 \sigma^{12/5}_R (d_{k}))^{5/6}} \right) - 1\right]^{-1},
\end{equation}
\end{subequations}where $\sigma^2_R(d_k) = 1.23 C^2_n w^{7/6} d_k^{11/6}$ is the Rytov variance~\cite{Andrews2005}, $w$ is the wave number\footnote{The wave number is defined as the spatial frequency of a wave and, in this work, we use it as the number of radians per unit distance.} and $C^2_n = 1.7 \times 10^{-14}~\text{m}^{-2/3}$ denotes the refractive index structure parameter, which is used to characterize the atmospheric turbulence.
From~\cite{Farid.2007}, the pointing loss is given by $I_{p,E}~=~\exp(-\frac{2 \tau ^2}{\omega _e^2})$, with $\omega _e~=~(\sqrt{\pi } \omega _b^2 \erf(\nu ) / (2 \nu  \exp \left(-\nu ^2\right)))^{1/2}$ as the equivalent beamwaist and $\tau$ as the radial displacement at the receiver. Considering that Eve's displacement follows an independent and identical Gaussian distribution with standard deviation $\sigma_s$ for both vertical and horizontal axis, the pdf of the pointing errors can be then expressed as~\cite{Sandalidis.2009}
\begin{equation}
\label{eq:perrorpdf}
f^p(I_{p,E})~=~\xi ^2 I_{p,E}^{\xi ^2-1},
\end{equation}where $\xi~=~\omega_e / (2 \sigma_s)$. The SNR from~\eqref{eq1} can be rewritten as
\begin{equation}
\gamma_k = A_0 \gamma_0 \mathcal{I}_{k} = \frac{A_0 \mathcal{I}_k}{N_0},
\label{eqGammak1}
\end{equation}where $\gamma_{0} = \frac{1}{N_0} = \frac{\eta_e E_s A}{h f_o \Delta_f}$, is the turbulence and pointing error free SNR which does not take into account $\mathcal{I}_k$ and the fraction of the available power $A_0$ received by node $k$. Note that the worst case of a pointing errors-free eavesdropper can be obtained by setting $\sigma_s=0$ in~\eqref{eq:perrorpdf}.

Since the laser beam emitted by Alice suffers divergence due to optical diffraction, following the approach described in~\cite{Martinez2015}, we assume that Eve is located in the divergence region, implying that Eve is close to Bob~\cite{Andrews1999} and is able to obtain part of the laser beam not captured by Bob, as shown in Fig.~\ref{fig:figMS}. In such approach, communication is inherently secure for small divergence angles but, for long distances, Eve has a better chance to eavesdrop.

From~\cite{Gappmair.2010}, we have that when both pointing error and atmospheric turbulence are considered, i.e., $\mathcal{I}_{E}^{i,j}~=~I_{a,E}^{i,j}~I_{p,E}^{i,j}$, the irradiance distribution is given by
\begin{equation}
\label{eq:pdfggp}
f^{\gamma\gamma p}(I)~=~\frac{\left(\alpha  \beta  \xi ^2\right) G_{1,3}^{3,0}\left((\alpha  \beta ) I\left|
\begin{array}{c}
 \xi ^2 \\
 \xi ^2-1,\alpha -1,\beta -1 \\
\end{array}
\right.\right)}{\Gamma (\alpha ) \Gamma (\beta )},
\end{equation}where $G_{m,n}^{p,q}\left(\cdot\right)$ is the Meijer-G function.
Finally, it is worthy mentioning that we assume the phase distortion to be negligible, which can be achieved in practice through the use of modal compensation techniques such as, e.g., Zernike polynomials~\cite{Noll.76}.

\subsection{Transmission Schemes}

The MIMOME FSO scenario adopted in this work employs two transmission techniques from~\cite{yan.2015}, namely adaptive and fixed-rate transmission schemes. %\commbr{In order to use such schemes, Alice must estimate the average CSI of the eavesdropper channel, which is an assumption commonly adopted in the literature (see, for instance,~\cite{yan.14.sisome,yan.2015,Martinez2015}). Noting that Eve is close to Bob, the path-loss, large-scale and small-scale parameters experienced by Bob and Eve can be assumed the same~\cite{Martinez2015}. However, parameters related to pointing errors (i.e., standard deviation of pointing error displacement $\sigma_s$) are still unknown such that, for a given practical scenario, the system can be projected to be secure for a worst-case value of $\sigma_s$.}
Then, in order to asses the performance of such schemes, the average CSI of the eavesdropper channel must be assumed at Alice, which is the strategy commonly adopted by the literature (c.f.,~\cite{yan.14.sisome,yan.2015,Martinez2015}). Noting that Eve is usually close to Bob in these scenarios, large-scale and small-scale parameters experienced by Bob and Eve can be assumed similar~\cite{Martinez2015}. Therefore, although the average CSI with respect to Eve is very unlikely to be known in practice, the system can still be designed to be secure for a worst-case scenario.

%Note that the performance of both schemes rely on the assumption that Alice possess only the average CSI with respect to Eve, without knowing its instantaneous CSI. This is an assumption commonly adopted in the literature (see, for instance,~\cite{yan.14.sisome,yan.2015,Martinez2015}) and supported by the fact that, noting that Eve is close to Bob, the path-loss, large-scale and small-scale parameters experienced by Bob and Eve can be assumed the same~\cite{Martinez2015}, which gives statistical CSI knowledge about Eve.

\subsubsection{Adaptive Transmission Scheme}
In the case when Alice possesses instantaneous CSI with respect to Bob, and average SNR about Eve, 
%This scheme assumes that Alice possesses instantaneous CSI with respect to Bob, having only the average \commbr{estimated} SNR about Eve. Thus, 
Alice is able to calculate the instantaneous channel capacity $C_B$ and consequently adjust $\mathcal{R}_E$ and $\mathcal{R}$ according to $C_B$, subject to the constraint $ 0 \leq \mathcal{R}_B \leq C_B$. This guarantees the reliability constraint. In turn, the violation of the secrecy constraint is defined as the probability that the equivocation rate $\mathcal{R}_E$ is less than the capacity of the eavesdropper channel ($C_E$).

\subsubsection{Fixed-Rate Transmission Scheme}
%%#############################################################################
Differently from the adaptive scheme, for the fixed-rate transmission scheme both instantaneous $C_B$ and $C_E$ are unavailable at Alice, meaning that Alice has only the average SNR of the main and eavesdropper's channels. This scheme is more practical (and less complex) than the Adaptive scheme since it requires no feedback from Bob to Alice. Note also that, in this scheme, both the reliability and the secrecy constraints are not guaranteed, and one must determine $R_B$ and $R_E$ that jointly maximize the EST.

%
%Note that the performance of both schemes rely on the assumption that Alice possess only the average CSI with respect to Eve, not having any information about its instantaneous CSI. This is an assumption commonly adopted in the literature  and supported by the fact that the turbulence and pointing error-free average value can be obtained directly from~\eqref{eqGammak1}.
%

Note that the performance of both schemes rely on the assumption that Alice possess the average CSI with respect to Eve, not having any information about its instantaneous CSI. This is an assumption commonly adopted in the literature and supported by the fact that a worst-case scenario can be established based on large-scale and small-scale parameters experienced by Bob (assumed to be similar to Eve), while the turbulence and pointing error-free average value can be obtained directly from~\eqref{eqGammak1}.

%\section{Performance Analysis for MIMOME FSO Communications} \label{sec:mimome_fso}

\subsection{Effective Secrecy Throughput with Eavesdropper Outage Constraints}

The EST is a metric proposed in~\cite{yan.14.sisome} that uses both  reliability and secrecy constraints to determine the throughput of the wiretap channel, being defined as~\cite{yan.2015}
\begin{equation} \label{eq:effective_secrecy_throughput}
\Psi(\mathcal{R}_E, \mathcal{R}_B) =  (\mathcal{R}_B - \mathcal{R}_E) \overbrace{\left[1-\mathcal{T}(\mathcal{R}_B)\right]}^{\text{reliability}} \overbrace{\left[1-\mathcal{S}(\mathcal{R}_E)\right]}^{\text{secrecy}},
\end{equation}where $\mathcal{T}(\mathcal{R}_B) = \Pr\{\mathcal{R}_B > C_B\}$ corresponds to the outage probability (in terms of reliability) and $\mathcal{S}(\mathcal{R}_E) = \Pr\{\mathcal{R}_E \leq C_E\}$ represents the SOP.

Even though the EST presented in~\cite{yan.14.sisome} is a useful performance metric, it does not impose any constraint regarding the SOP, which means that Eve might operate at a very low outage probability, acquiring a confidential information and, thus, compromising secrecy. In~\cite{Monteiro2015}, the authors circumvent this problem by imposing a constraint and defining the EST as
\begin{equation} \label{eq:effective_secrecy_throughput2}
\begin{split}
\Psi^{\textsf m}(\mathcal{R}_E, \mathcal{R}_B) = \left\{
\begin{array}{ll}
  \Psi(\mathcal{R}_E, \mathcal{R}_B), &  \text{if } \mathcal{S}(\mathcal{R}_E) \leq \mathcal{S}^{\text{th}};  \\[5pt]
  0, &  \text{if } \mathcal{S}(\mathcal{R}_E) > \mathcal{S}^{\text{th}},
\end{array}
\right.
\end{split}
\end{equation}where $\Psi(\mathcal{R}_E, \mathcal{R}_B)$ and $\mathcal{S}(\mathcal{R}_E)$ represent respectively the constrained EST and the SOP, and $\mathcal{S}^{\text{th}}$ is the maximum allowed value of $\mathcal{S}(\mathcal{R}_E)$. For simplicity, in the rest of this work we drop the index ${\textsf m}$ in the EST from~\eqref{eq:effective_secrecy_throughput2} and consider the EST with no constraints as the particular case where $\mathcal{S}^{\text{th}} = 1$.

\section{EST of Adaptive MIMOME FSO Communication} \label{sec:mimome_fso_adap}

Since Alice possesses the instantaneous CSI regarding the legitimate channel, the reliability constraint is always guaranteed (i.e., $\mathcal{T}^{a}(\mathcal{R}_B) = 0$) in the adaptive scheme. The EST from~\eqref{eq:effective_secrecy_throughput} can then be adjusted to the adaptive scheme as:
\begin{equation} \label{eq:effective_secrecy_throughput_a}
\Psi^{a}(\mathcal{R}_E) = (C_B - \mathcal{R}_E)\, \left[ 1 - \mathcal{S}(\mathcal{R}_E)\right].
\end{equation}

In order to obtain a closed-form expression to the EST of the adaptive scheme, one needs to evaluate the SOP:
\begin{equation} \label{eq:prsop}
\begin{split}
\mathcal{S}(\mathcal{R}_E) & = \Pr\{C_E > \mathcal{R}_E\} = \Pr\left\{\gamma_{E} > 2^{\mathcal{R}_E}-1\right\}\\
&= \Pr\left\{ \frac{\mathcal{I}_E A_0}{N_0} > 2^{\mathcal{R}_E}-1 \right\}.
\end{split}
\end{equation}
Note that solving~\eqref{eq:prsop} is not a straightforward task since the irradiance is composed of the turbulence (which is modeled as the summation of random variables due to the diversity combining technique), and another random variable that represents the pointing errors.
\begin{lemma}\label{lemma1}
The SOP of the adaptive scheme\footnote{The SOP presented in Lemma~\ref{lemma1} is also valid for the fixed-rate scheme since only the average SNR is assumed at Alice for both schemes.}  is given by
\begin{equation} \label{eq:sop}
\mathcal{S}(\mathcal{R}_E) = 1 - F_{E}^{\gamma\gamma p}(\mathcal{X}_E),
\end{equation}where $\mathcal{X}_E = \frac{N_0\left(2^{\mathcal{R}_E}-1\right)}{N_EA_0}$, and $F_{E}^{\gamma\gamma p}(\cdot)$ represents the cdf  when both pointing errors and atmospheric turbulence are considered, which is given by
\begin{equation} \label{eq:CDFGammaGammaP}
\begin{split}
&F_{E}^{\gamma\gamma p}(\mathcal{X}_{E}) = \frac{\pi }{\Gamma \left(\alpha _E\right) \Gamma \left(\beta _E\right)}   \bigg( -\csc \left(\pi  \left(\alpha _E-\beta _E\right)\right) \\
&\!\left.\sum _{u=1}^2 \sum _{v=1}^2 {c}_u {c}_v \mathcal{X}_E^{{b}_u}\!\!\left(\alpha _E \beta _E\right)^{{b}_u}\!\left( \Gamma \left({a}_v\right)\! _1\tilde{F}_2\left({a}_v;{d}_v,{e}_v;\mathcal{X}_E \alpha _E \beta _E\right)\right) \right.\\
&\left. + \frac{\pi  \mathcal{X}_E^{\xi ^2} \left(\alpha _E \beta _E\right)^{\xi ^2} \csc \left(\pi  \left(\alpha _E-\xi ^2\right)\right) \csc \left(\pi  \left(\beta _E-\xi ^2\right)\right)}{\Gamma \left(\xi ^2-\alpha _E+1\right) \Gamma \left(\xi ^2-\beta _E+1\right)}\right),
\end{split}
\end{equation}where $\alpha_{k}=\alpha(d_k)$ and $\beta_{k}=\beta(d_k) N_k$ represent, respectively, the large-scale and the small-scale parameters related to the number of cells in the scattering process~\cite{Habash2001}, $_1\tilde{F}_2(\cdot)$ denotes the regularized hypergeometric function, and $a_x,~x~\in~\{u,v\}$ represents the $x$-th element of vector $\bm{a} = \left[b_u,b_u-\xi ^2\right]$, which is also valid for vectors $\bm{b} = \left[\alpha,\beta\right]$, $\bm{c} = [-1, 1]$, $\bm{d} = \left[b_u+1,\left(\beta-\alpha\right) c_u+1\right]$ and $\bm{e} = \left[\left(\beta-\alpha\right) c_u+1, b_u-\xi ^2+1\right]$.
\end{lemma}
\begin{IEEEproof}
Please refer to Appendix~\ref{appendix:lemma1}.
\end{IEEEproof}

The EST is then obtained by placing~\eqref{eq:sop} in~\eqref{eq:effective_secrecy_throughput_a}.

\subsection{Optimal Target Secrecy Rate}
%$\mathcal{R}^{\star}_{a},$\mathcal{R}^{\star}_{o}$

In order to maximize the EST, the optimal value of the secrecy rate $\mathcal{R}$  must be obtained. Noting that $\mathcal{R}~=~\mathcal{R}_B - \mathcal{R}_E$ and, for the adaptive scheme, Alice has the instantaneous capacity of the legitimate channel  such that $\mathcal{R}_B~=~C_B$, following~\cite{yan.2015} we choose to obtain the optimal value of the rate of redundancy $\mathcal{R}_E$, which can be used directly to obtain $\mathcal{R}$. When evaluating the EST from~\eqref{eq:effective_secrecy_throughput_a}, one can see that while a larger $\mathcal{R}_E$ leads to a smaller value of $(C_B - \mathcal{R}_E)$, it simultaneously increases $1 - \Pr\{C_E > \mathcal{R}_E\}$. Thus, one could expect to exist an optimal value of $\mathcal{R}_E$ that maximizes the EST. When considering an outage constrained scenario, however, one needs to check whether such optimal value meets the outage constraint or not. In this sense, we have the following result.

\begin{theorem}\label{theo1}
The value of $\mathcal{R}_E$ that maximizes the outage-constrained EST for the MIMOME FSO adaptive scheme is% obtained as
\begin{equation} \label{eq:optimal_r_a}
\mathcal{R}_{E}^{a^\star} = \max\left(\mathcal{R}^{a^\star}_{E,u},\mathcal{R}^{{th}^\star}_{E}\right),
\end{equation}
where $\mathcal{R}^{a^\star}_{E,u}$ is the unconstrained optimal value of $\mathcal{R}_E$ given by the solution of the fixed-point equation 
%\begin{subequations}
\begin{equation} \label{eq:theo1}
\begin{split}
&\mathcal{R}^{a^\star}_{E,u} = (C_B-C_B 2^{R_{E,u}^{a^*}}+2^{R_{E,u}^{a^*}} R_{E,u}^{a^*})+\\
&\frac{4 \sigma _s^2 \left(2^{R_{E,u}^{a^*}}-1\right)^2}{\log (2) \omega _E^2 2^{R_{E,u}^{a^*}}}+\frac{A_0 N_E \theta _E^{\text{AP}}}{N_0 \log (2) \omega _E^2 2^{R_{E,u}^{a^*}} \text{E}_{\vartheta-1}\left(\frac{\mathcal{X}_E}{\theta _E^{\text{AP}}}\right)}\\
&\left(\frac{\left\{2^{R_{E,u}^{a^*}} \left[\log (2) \omega _E^2 \left(C_B-R_{E,u}^{a^*}\right)-4 \sigma _s^2\right]+4 \sigma _s^2\right\}}{\exp\left({\frac{\mathcal{X}_E}{\theta _E^{\text{AP}}}}\right)}\right.\\
&\left.-\frac{\left(\frac{\mathcal{X}_E}{\theta _E^{\text{AP}}}\right)^{-k} \left(\omega _E^2-4 k_E^{\text{AP}} \sigma _s^2\right)\left[\Gamma \left(k_E^{\text{AP}}\right)-\Gamma \left(k_E^{\text{AP}},\frac{\mathcal{X}_E}{\theta _E^{\text{AP}}}\right)\right]}{\left(2^{R_{E,u}^{a^*}}-1\right)^{-1}}\right),
\end{split}
\end{equation}and $\mathcal{R}^{{th}^\star}_{E}$ is the constrained optimal value of $\mathcal{R}_{E}$ for a given maximum allowed $S^{\text{th}}$, which is given by
\begin{equation}\label{eq:optimal_rOeth_a}
\begin{split}
&\mathcal{R}^{{th}^\star}_{E} = \\
&\log_2\left(1 + \frac{A_0 N_E \theta_E^{\text{AP}}}{N_0} \left(\frac{\Gamma\left(k_E^{\text{AP}},\frac{\mathcal{X}_E}{\theta _E^{\text{AP}}}\right) - S^{\text{th}} \Gamma\left(k_E^{\text{AP}}\right)}{\text{E}_{\vartheta}\left(\frac{\mathcal{X}_E}{\theta _E^{\text{AP}}}\right)}\right)^{\frac{1}{k_E^{\text{AP}}}}\right).
\end{split}
\end{equation}
%\end{subequations}}}
%\begin{equation} \label{eq:theo1}
%\begin{split}
%\mathcal{R}^{a^\star}_{E,u}& = C_B + \\
%& \frac{W\left(-\frac{N_E A_0^2 2^{1-C_B} \mathcal{X}_E^2 \Gamma \left(k_E^{\text{AP}}\right) e^{\frac{\mathcal{X}_E}{\theta_E^{\text{AP}}}} \left(\frac{\mathcal{X}_E}{\theta_E^{\text{AP}}}\right)^{-k_E^{\text{AP}}} C_2}{N_0}\right)}{\log(2)} ,\\
%\end{split}
%\end{equation}
%$W(\cdot)$ corresponds to the Lambert $W$-function, and

In~\eqref{eq:theo1} and~\eqref{eq:optimal_rOeth_a}, $\text{E}_{\cdot}(\cdot)$ is the exponential integral function, $\vartheta~=~-k_E^{\text{AP}}+\frac{\omega _E^2}{4 \sigma _s^2}+1$, $\Gamma(\cdot,\cdot)$ is the incomplete gamma function, while $\theta_k^{\text{AP}}$ and $k_k^{\text{AP}}$ are the scale and shape parameters of the approximated gamma variable\footnote{Note that, to obtain~\eqref{eq:theo1}, the gamma-gamma variable in~\eqref{eq:prsop} is approximated as a gamma variable as described in Appendix~\ref{appendix:theo1}.}, which are respectively given by
\begin{subequations}
\begin{eqnarray}\label{eq:kapk}
& k^{\text{AP}}_k = \left[\frac{(\beta_k +1)(\alpha_k  + 1)}{\beta_k  \alpha_k }-(1+\epsilon )\right]^{-1}, \\
\label{eq:thetaapk}
& \theta^{\text{AP}}_k = \left[\frac{(\beta_k +1)(\alpha_k +1)}{\beta_k  \alpha_k }-(1+\epsilon )\right] \Omega,
\end{eqnarray}
\end{subequations}where $\epsilon$ and $\Omega$ are adjustment parameters~\cite{Ahmadi2009}.
\end{theorem}
\begin{IEEEproof}
Please refer to Appendix~\ref{appendix:theo1}.
\end{IEEEproof}

\section{EST of Fixed-Rate MIMOME FSO Communication} \label{sec:mimome_fso_fixe}

In the fixed-rate scheme, both reliability and secrecy cannot be guaranteed, and the EST is obtained as
\begin{equation} \label{eq:effective_secrecy_throughput_f}
\Psi^{\textsf{f}}(\mathcal{R}_E, \mathcal{R}_B) =  (\mathcal{R}_B - \mathcal{R}_E) \left[1-\mathcal{T}^{f}(\mathcal{R}_B)\right] \left[1-\mathcal{S}(\mathcal{R}_E)\right].
\end{equation}
%where the superscript index $a$ and $f$ from, respectively, ~\eqref{eq:effective_secrecy_throughput_a} and~\eqref{eq:effective_secrecy_throughput_f}, are used to indicate the adaptive and fixed-rate schemes.

The SOP in~\eqref{eq:effective_secrecy_throughput_f} is obtained from~\eqref{eq:sop}. The reliability probability, in turn, is
\begin{equation} \label{eq:probTranFixed}
\begin{split}
1-\mathcal{T}^{f}(\mathcal{R}_B) &= \Pr \{ C_B > \mathcal{R}_B \}\\
&= \Pr \{ \gamma_{B} > 2^{\mathcal{R}_B} - 1 \}\\
&= \Pr\left\{ \frac{\mathcal{I}_B A_0}{N_0} > 2^{\mathcal{R}_B}-1 \right\}.
\end{split}
\end{equation}

Having in mind that the turbulence $\mathcal{I}_B$ in~\eqref{eq:probTranFixed} encompasses the effects of both TLS and MRC, one has that~\eqref{eq:probTranFixed} is given as follows.

\begin{lemma}\label{lemma2}
The outage probability in terms of reliability for the fixed-rate scheme is given by
\begin{equation} \label{eq:lemma2}
\mathcal{T}^{f}(\mathcal{R}_B) = F_{B}^{\gamma\gamma}(\mathcal{X}_B)^{N_A} ,
\end{equation}where $\mathcal{X}_B = \frac{N_0\left(2^{\mathcal{R}_B}-1\right)}{N_BA_0}$ and $F_{B}^{\gamma\gamma}(\cdot)$, represents the cdf of a single gamma-gamma random variable for the SNR for Bob,  and is given by~\cite{Habash2001}
\begin{equation} \label{eq:CDFGammaGamma}
\begin{split}
&F_{B}^{\gamma\gamma}(\mathcal{X}_{B}) = \frac{\pi }{\Gamma \left(\alpha_{B}\right) \Gamma \left(\beta_{B}\right) \sin \left(\pi  \left(\alpha_{B}-\beta_{B}\right)\right)}\\
& \left[ \frac{\left(\mathcal{X}_{B} \alpha_{B} \beta_{B} \right)^{\beta_{B}} \, _1F_2\left(\beta_{B};\beta_{B}+1,-\alpha_{B}+\beta_{B}+1;\alpha_{B} \beta_{B} \mathcal{X}_{B}\right)}{\beta_{B} \Gamma \left(-\alpha_{B}+\beta_{B}+1\right)} \right. \\
& \left. -\frac{\left(\mathcal{X}_{B} \alpha_{B} \beta_{B}\right)^{\alpha_{B}} \, _1F_2\left(\alpha_{B};\alpha_{B}+1,\alpha_{B}-\beta_{B}+1;\alpha_{B} \beta_{B} \mathcal{X}_{B}\right)}{\alpha_{B} \Gamma \left(\alpha_{B}-\beta_{B}+1\right)} \right],
\end{split}
\end{equation}where $_1F_2(\cdot)$ denotes the generalized hypergeometric function.
\end{lemma}
\begin{IEEEproof}
In order to obtain~\eqref{eq:lemma2}, the same approach used in Lemma~\ref{lemma1} has been adopted, with the difference that~\eqref{eq:lemma2} does not take into account the pointing errors and it presents the effect of TLS, noting that it is related to the legitimate channel.
\end{IEEEproof}

%Using a similar approach as presented in Appendix~\ref{appendix:lemma1},~\eqref{eq:probTranFixed} can be approximated as \commbr{Esse passo intermediário é necessário de ser apresentado aqui? Creio que~\eqref{eq:provaprtx2fixed} poderia ser movida para o apêndice também.}
%\begin{equation}\label{eq:provaprtx2fixed}
%\text{Pr}^f_{\text{tx}}(\mathcal{R}_B) \approx \Pr \left\{ X_{\alpha_B} Y_{\beta_B} > \sqrt{\frac{N_0\left(2^{\mathcal{R}_B}-1\right)}{N_B A_0^2} } \right\}^{N_A},
%\end{equation}
%where $X_{\alpha_B}~\sim~\Gamma \left( N_B \alpha, \frac{1}{N_B \alpha} \right)$ and $Y_{\beta_B}~\sim~\Gamma \left( N_B \beta, \frac{1}{N_B \beta} \right)$. Finally, the probability of transmission for the fixed-rate transmission scheme can be approximated, from~\eqref{eq:provaprtx2fixed}, by

The EST of the fixed-rate scheme is finally obtained after replacing~\eqref{eq:sop} and~\eqref{eq:lemma2} in~\eqref{eq:effective_secrecy_throughput_f}.

\subsection{Optimal Target Secrecy Rate }

Differently from the adaptive scheme, in the fixed-rate scheme the EST is a function of both $\mathcal{R}_E$ and $\mathcal{R}_B$. Thus, in order to obtain the optimal values of such parameters (i.e., $\mathcal{R}_{E}^{f^\star}$ and $\mathcal{R}_{B}^{f^\star}$), one must first identify the optimal values of  $\mathcal{R}_E$ and $\mathcal{R}_B$ without secrecy constraints, which is presented in what follows.% in line with the results from~\cite{yan.2015}.
\begin{lemma}\label{lemma3}
The unconstrained optimal values of $\mathcal{R}_E$ and $\mathcal{R}_B$ that jointly achieve a locally maximum EST are given, respectively, by 
\begin{subequations}
\begin{equation} \label{eq:lemma3eq1}
\begin{split}
&\mathcal{R}_{E,u}^{f^\star} = \mathcal{R}_{B,u}^{f^\star} + \frac{1}{\log (2) N_A} \times\\
&\left[\left(1-2^{-\mathcal{R}_B}\right) e^{\frac{\mathcal{X}_B}{\theta_B^{\text{AP}}}} \Gamma \left(k_B^{\text{AP}}\right) \left(\frac{\mathcal{X}_B}{\theta_B^{\text{AP}}}\right)^{-k_B^{\text{AP}}} \left({C_1}-{C_1}^{1-N_A}\right)\right],
\end{split}
\end{equation}
\begin{equation} \label{eq:lemma3eq2}
\begin{split}
&\mathcal{R}_{B,u}^{f^\star} = \mathcal{R}_{E,u}^{f^\star}+\frac{\left(-1+2^{R_{E,u}^{f*}}\right)}{2^{R_{E,u}^{f*}}\omega _e^2 \log (2)} \left(\frac{4 \sigma _s^2}{2^{-R_{E,u}^{f*}}\omega _e^2 \log (2)} + \right. \\
&\left.\frac{\left(\frac{X_{R_E}}{\theta _E^{\text{AP}}}\right)^{-k_E^{\text{AP}}} \left(\omega_e^2-4 k_E^{\text{AP}} \sigma _s^2\right) \left[\Gamma \left(k_E^{\text{AP}}\right)-\Gamma \left(k_E^{\text{AP}},\frac{X_{R_E}}{\theta _E^{\text{AP}}}\right)\right]}{-\frac{\left(-1+2^{R_{E,u}^{f*}}\right) E_{\vartheta-1}\left(\frac{X_{R_E}}{\theta _E^{\text{AP}}}\right) N_0}{\theta _E^{\text{AP}} A_0 N_E}+\exp\left(-\frac{X_{R_E}}{\theta _E^{\text{AP}}}\right)}\right),
\end{split}
\end{equation}
\end{subequations}where $C_1 = Q\left(k_B^{\text{AP}},0,\frac{\mathcal{X}_B}{\theta_B^{\text{AP}}}\right)$, being $Q(\cdot,0,\cdot)$ the generalized regularized incomplete gamma function.
% and $\mathcal{X}_B$ is defined as
%\begin{equation}
%\mathcal{X}_B=\sqrt{\frac{\left(2^{\mathcal{R}_B}-1\right) N_0}{N_B A_0^2}}.
%\end{equation}
\end{lemma}
\begin{IEEEproof}
Please refer to Appendix~\ref{appendix:lemma3}.
\end{IEEEproof}

The optimal values of $\mathcal{R}_E$ and $\mathcal{R}_B$ for the EST with secrecy constraints are then presented in whats follows.

\begin{theorem}\label{theo2}
The optimal constrained values of $\mathcal{R}_E$ and $\mathcal{R}_B$ that maximize the EST with secrecy constraints for the fixed-rate scheme are given, respectively, by
\begin{subequations}
\begin{equation} \label{eq:theo2eq1}
\mathcal{R}_{E}^{f^\star} = \max\left(\mathcal{R}_{E,u}^{f^\star},\mathcal{R}^{{th}^\star}_{E}\right),
\end{equation}
\begin{equation} \label{eq:theo2eq2}
\begin{split}
\mathcal{R}_{B}^{f^\star} = \left\{
\begin{array}{ll}
  \mathcal{R}_{B,u}^{f^\star}, &  \text{if } \mathcal{R}_{E,u} \geq \mathcal{R}^{{th}^\star}_{E};  \\[5pt]
  \mathcal{R}_{B,c}^{f^\star}, &  \text{otherwise}.
\end{array}
\right.
\end{split}
\end{equation}
\end{subequations}where $\mathcal{R}_{B,c}^{f^\star}$ is the constrained optimal value of $\mathcal{R}_B$ and is given by
%Por alguns testes que fiz, a outra função (usando o productlog) funcinou melhor.
%\begin{equation} \label{eq:theo2eq3}
%\begin{split}
%\mathcal{R}_{B,c}^{f^\star} = \mathcal{R}_E + \frac{\theta_B^{\text{AP}} N_b r_b^2 \left(C_1^{1-N_A}-C_1\right) \Gamma \left(k_B^{\text{AP}}\right) \mathcal{X}_B}{N_0 2^{\mathcal{R}_B-1} \log (2) N_A e^{-\frac{\mathcal{X}_B}{\theta_B^{\text{AP}}}} \left(\frac{\mathcal{X}_B}{\theta_B^{\text{AP}}}\right)^{k_B^{\text{AP}}-1}}.
%\end{split}
%\end{equation}
\begin{equation} \label{eq:theo2eq3}
\begin{split}
&\mathcal{R}_{B,c}^{f^\star} = \log_2 \Bigg(-\frac{A_0 N_B \theta_B^{\text{AP}}}{N_0} \\
&\left. \text{W}\left(\frac{\left(\text{C1}-\text{C1}^{1-N_A}\right) \Gamma \left(k_B^{\text{AP}}\right) \left(\frac{X_{R_B}}{\theta _B^{\text{AP}}}\right)^{1-k_B^{\text{AP}}}}{\exp\left(\frac{N_0}{A_0 N_B \theta _B^{\text{AP}}}\right) \left(R_{B,c}^{f*}-R_E^{f*}\right)\log (2)N_A}\right)\right),\\
\end{split}%Tentei fazer a mudança que o João pediu, mas o tamanho dos termos da equação aumentam.
\end{equation}where $\text{W}(\cdot)$ corresponds to the Lambert $W$-function.
\end{theorem}
\begin{IEEEproof}
Please refer to Appendix~\ref{appendix:theo2}.
\end{IEEEproof}

\begin{figure} [!t]
	\centering
	\includegraphics[width=240px]{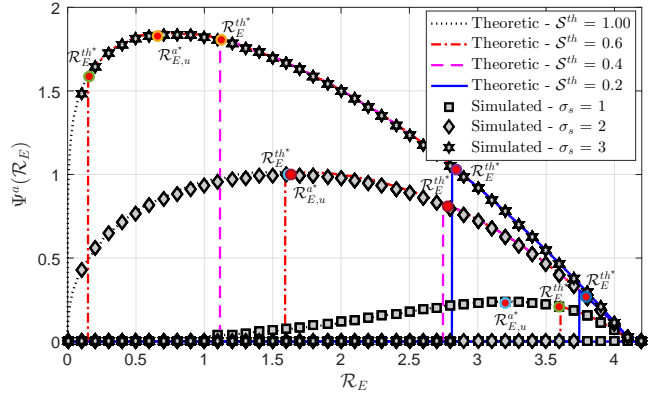}
	\caption{$\Psi^{a}(\mathcal{R}_E)$ versus $\mathcal{R}_E$ for adaptive transmission scheme with $N_A~=~N_E~=~2$, $N_B~=~1$, $\mathcal{S}^{\text{th}} \in \{1,0.6,0.4,0.2\}$ and $\sigma_s~\in~\{1,2,3\}$.}
	\label{fig:EST_RE_Adaptive}
\end{figure}

%%#############################################################################
\section{Numerical results} \label{sec:numerical_results}

%\textbf{\footnote{{\bf Note that increasing the number of receive apertures at Bob has the same effect as increasing the number of receive apertures at Alice.}}}
In this section, we present some numerical results in order to evaluate the previous analysis, adopting the same turbulence-free SNR $\gamma_{0}$ for both Eve and Bob, with $d_k=d_B=d_E=1$~km, $\lambda=1550~$nm~\cite{Rjeily2015}, and using the adjustment $\epsilon=0$ and $\Omega=0.97$ for the approximated gamma variable\footnote{Using the approximation proposed in~\cite{Ahmadi2009}, $\epsilon$ and $\Omega$ must be chosen in order to minimize the difference between the results obtained by the cdf of gamma-gamma and gamma random variables.}, unless stated otherwise. Following~\cite{Farid.2007}, we also use $\omega_b~=~2.5$ and $\rho~=~0.1$.

Fig.~\ref{fig:EST_RE_Adaptive} presents the EST with secrecy constraints versus the redundancy rate for the adaptive transmission scheme with $\sigma_s~\in~\{1,2,3\}$ and $N_A=N_E=2$, for different values of $\mathcal{S}^{\text{th}}$. One can see that, as the standard deviation of pointing error displacement increases, the maximum allowed SOP is also increased, meaning that, depending on $\sigma_s$, the system might have to operate with a lower value of $\Psi^{a}(\mathcal{R}_E)$ in order to ensure that the maximum allowed SOP is feasible. Note that this is in accordance with the proposed system model, since the increase of $\sigma_s$ decreases the fraction of the power received by Eve, provided that it increases the probability of the eavesdropper being outside the received beam radius $\omega_b$. From Fig.~\ref{fig:EST_RE_Adaptive}, we can also see that the approximated theoretic values (represented by the red circles) of $\mathcal{R}^{a^\star}_{E,u}$ (unconstrained) and $\mathcal{R}^{{th}^\star}_{E}$ (for $\mathcal{S}^{\text{th}} \in \{0.6,0.4,0.2\}$) from, respectively, \eqref{eq:theo1} and~\eqref{eq:optimal_rOeth_a}, and represented by red dots are in good agreement with the optimal numerical results for different values of $\sigma_s$, demonstrating an approximation error below $2\%$. Note also that, as stated in Appendix~\ref{appendix:theo1} and similar to that seen in~\cite{yan.14.sisome,yan.2015,Monteiro2015}, the stationary points obtained from~\eqref{eq:theo1} represent the local maximum for all the scenarios evaluated in this work.
 %, more importantly, the approximation error of the optimal $\Psi^{a}(\mathcal{R}^{a^\star}_{E,u})$ is only about $1\%$ from the numeric value.

\begin{figure}[!t]
	\centering
	\includegraphics[width=240px]{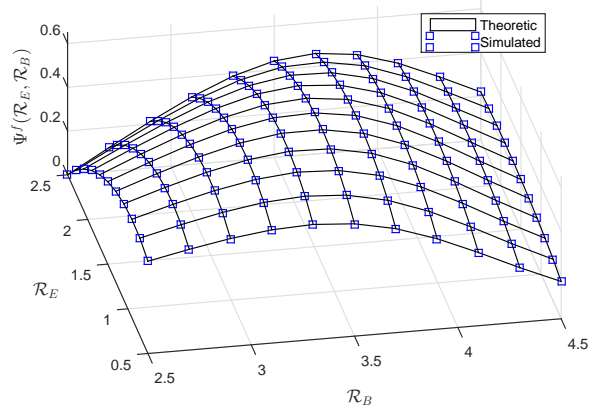}
	\caption{$\Psi^{f}(\mathcal{R}_E, \mathcal{R}_B)$ versus $\mathcal{R}_E$, $\mathcal{R}_B$ for fixed-rate transmission scheme with $N_A~=~N_E~=~2$, $N_B~=~1$, $\sigma_s~=~2$ and $\mathcal{S}^{\text{th}} = 1.0$.}
	\label{fig:Est_RE_RB_FixedRate_Aprox}
\end{figure}

In order to validate the analytical derivations of Lemmas~\ref{lemma1} and~\ref{lemma2}, Fig.~\ref{fig:Est_RE_RB_FixedRate_Aprox} presents the EST versus the redundancy rate $\mathcal{R}_E$ and the rate of transmitted codewords $\mathcal{R}_B$ for  $N_A=N_E=~2$, $N_B=1$ and $\sigma_s=2$ in an unconstrained scenario ($\mathcal{S}^{\text{th}} = 1.0$). We can see that the results using~\eqref{eq:sop} and~\eqref{eq:lemma2} match exactly the simulation results, confirming the usefulness of such equations. Moreover, note that the SOP derivation in~\eqref{eq:sop} is applied for both adaptive and fixed-rate schemes\footnote{While the EST obtained for the fixed-rate scheme requires both the SOP and the reliability probability, for the adaptive scheme only the SOP is required.}, such that Fig.~\ref{fig:Est_RE_RB_FixedRate_Aprox} also validates the obtained SOP expression for the adaptive scheme.

\begin{figure*}[!t]
	\centering
		\subfigure[Unconstrained scenario with $\mathcal{S}^{\text{th}} = 1.0$.]{\label{fig:Est_Re_Rb_FixedRate_nB1}
			\includegraphics[width=0.5\textwidth]{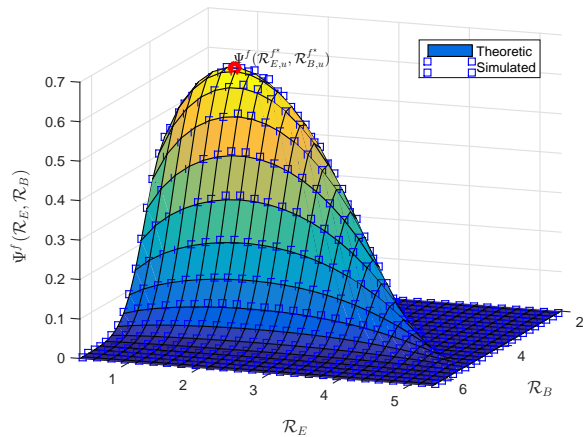}}
		\subfigure[Constrained scenario with $\mathcal{S}^{\text{th}} = 0.5$.]{\label{fig:Est_Re_Rb_FixedRate_Sth5}
			\includegraphics[width=0.5\textwidth]{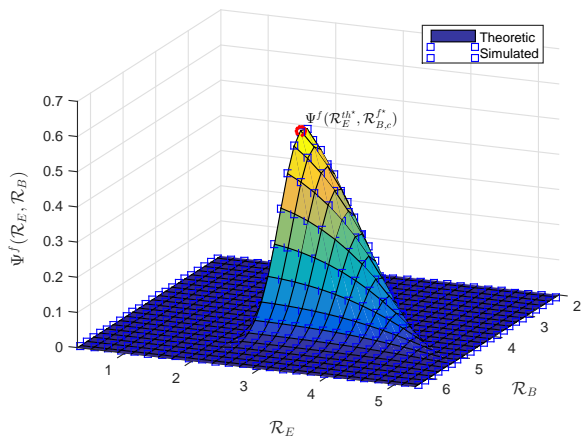}}
		\hfill
	\caption{$\Psi^{f}(\mathcal{R}_E, \mathcal{R}_B)$ versus $\mathcal{R}_E$, $\mathcal{R}_B$ for fixed-rate transmission scheme with $N_A~=~N_E~=~2$, $N_B~=~1$ and  $\sigma_s~=~2$.}
\end{figure*}

Fig.~\ref{fig:Est_Re_Rb_FixedRate_nB1} presents the unconstrained value ($\mathcal{S}^{\text{th}} = 1.0$) of the EST versus the redundancy rate $\mathcal{R}_E$ and the rate of transmitted codewords $\mathcal{R}_B$, for the fixed-rate transmission scheme with $N_A=N_E=2$, $N_B=1$ and $\sigma_s=2$. One can see that there is an optimal value of $\mathcal{R}_B$ for each value of $\mathcal{R}_E$ (and vice versa), and that there is a stationary point of $\Psi^{f}(\mathcal{R}_E, \mathcal{R}_B)$ that results in the optimal EST, as stated in Appendix~\ref{appendix:theo2}. We can also see that the approximated theoretic values of $\mathcal{R}^{f^\star}_{E,u}=1.250$~bpcu and $\mathcal{R}^{f^\star}_{B,u}=3.401$~bpcu from, respectively, \eqref{eq:lemma3eq1} and~\eqref{eq:lemma3eq2}, are in good agreement with the optimal numerical rates $\mathcal{R}_E = 1.257~$bpcu and $\mathcal{R}_B = 3.400$~bpcu, which results in $\Psi^{f}(\mathcal{R}_E, \mathcal{R}_B) = 0.621~$bpcu.

In Fig.~\ref{fig:Est_Re_Rb_FixedRate_Sth5} we present a similar analysis, but imposing a secrecy constraint $\mathcal{S}^{\text{th}} = 0.5$. The threshold value $\mathcal{R}^{{th}^\star}_{E}$, for which any value lower than that will result in an SOP greater than the threshold $\mathcal{S}^{th}$, can be obtained directly from~\eqref{eq:optimal_rOeth_a}. Note that, in agreement to Theorem~\ref{theo2}, the optimal value of $\mathcal{R}_{E}^{f^\star}$ is the maximum between $\mathcal{R}_{E,u}^{f^\star}$ and $\mathcal{R}^{{th}^\star}_{E}$, and that the optimal value $\mathcal{R}_{B}^{f^\star}$ can be obtained from~\eqref{eq:theo2eq3}. Finally, the optimal value of the $\Psi^{f}(\mathcal{R}_{E}^{th^\star}, \mathcal{R}_{B,c}^{f^\star})$  is presented using~\eqref{eq:effective_secrecy_throughput_f},~\eqref{eq:optimal_rOeth_a}~and~\eqref{eq:theo2eq3}, confirming the accuracy of the mathematical derivations.

\begin{figure}[!t]
	\centering
	\includegraphics[width=240px]{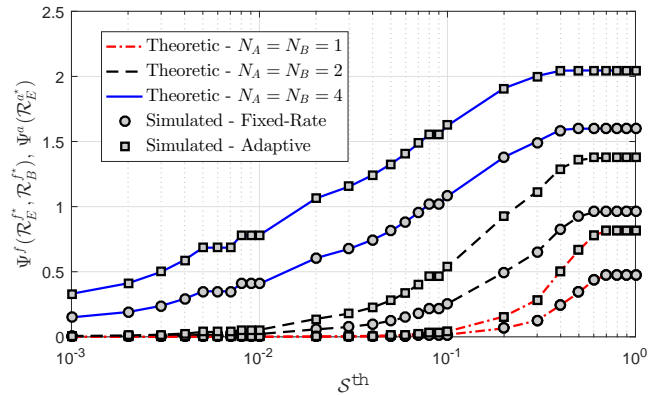}
	\caption{$\Psi^f(\mathcal{R}^{f^\star}_{E},\mathcal{R}^{f^\star}_{B})$, $\Psi^a(\mathcal{R}^{a^\star}_{E})$ versus $\mathcal{S}^{\text{th}}$ for fixed-rate and adaptive transmission schemes with $N_E~=~2$, $N_A~=~N_B \in \{1,2,4\}$ and $\sigma_s~=~2$.}
	\label{fig:Est_Sth_Adaptive_FixedRate}
\end{figure}

%\begin{figure}[!t]
%\centering
%\includegraphics[width=0.5\textwidth]{Est_Re_Rb_FixedRate_Sth5.eps}
%\caption{$\Psi^{f}(\mathcal{R}_E, \mathcal{R}_B)$ versus $\mathcal{R}_E$, $\mathcal{R}_B$ for the fixed-rate transmission scheme with $N_A~=~N_E~=~2$, $N_B~=~1$, $r_E~=~0.1$, $\mathcal{S}^{\text{th}} = 0.1$.}
%\label{fig:Est_Re_Rb_FixedRate_Sth5}
%\end{figure}
%Similar to that obtained in Fig.~\ref{fig:Est_RE_RB_FixedRate_Aprox}, in Fig.\eqref{fig:EST_RE_Adaptive_Aprox} we present the EST versus the redundancy rate $\mathcal{R}_E$. We can see that, similar to that obtained in the fixed-rate scenario, the approximation using the gamma CDF is close to that obtained using both~\eqref{eq:effective_secrecy_throughput_a} and the simulation results.
%
%\begin{figure}[!t]
%\centering
%\includegraphics[width=0.5\textwidth]{EST_RE_Adaptive_Aprox.eps}
%\caption{$\Psi^{a}(\mathcal{R}_E)$ versus $\mathcal{R}_E$ for the adaptive transmission scheme with $N_A~=~4$, $N_B~=~N_E~=~2$, $r_E~=~0.1$.}
%\label{fig:EST_RE_Adaptive_Aprox}
%\end{figure}
In Fig.~\ref{fig:Est_Sth_Adaptive_FixedRate} we present the EST versus $\mathcal{S}^{\text{th}}$ for the adaptive and fixed-rate schemes for $N_E=2$, $N_A=N_B \in \{1,2,4\}$ and $\sigma_s=2$. One can see that, for different values of $(N_A,N_B)$, the results using~\eqref{eq:effective_secrecy_throughput2},~\eqref{eq:effective_secrecy_throughput_a} and~\eqref{eq:effective_secrecy_throughput_f} are in perfect agreement with simulations. We can also see that, as the maximum allowed SOP increases, the maximum EST obtained by both schemes also increases, and that a higher number of apertures in the legitimate channel allows the system to achieve a lower SOP even in the unconstrained scenario. Furthermore, it is shown that the adaptive scheme is able to obtain a higher EST than that obtained using the fixed-rate scheme, which is expected since, when using the adaptive scheme, Alice has the instantaneous CSI about the legitimate channel.

\begin{figure}[!t]
	\centering
	\includegraphics[width=240px]{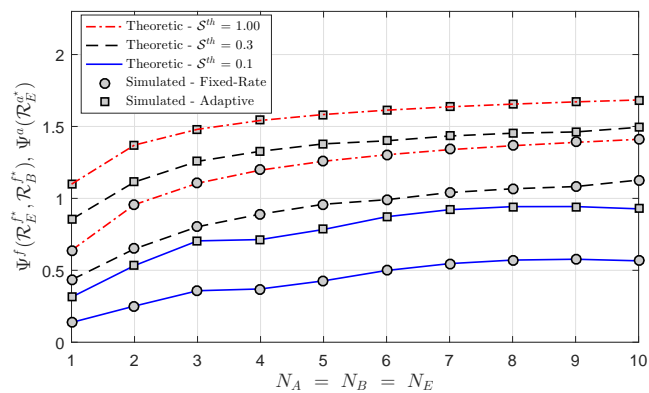}
	\caption{$\Psi^f(\mathcal{R}^{f^\star}_{E},\mathcal{R}^{f^\star}_{B})$, $\Psi^a(\mathcal{R}^{a^\star}_{E})$ versus $N_A~=~N_B~=~N_E$ for fixed-rate and adaptive transmission schemes with $\mathcal{S}^{\text{th}} \in \{1,0.3,0.1\}$.}
	\label{fig:Est_N_Adaptive_FixedRate}
\end{figure}

Fig.~\ref{fig:Est_N_Adaptive_FixedRate} presents the EST versus $N_A=N_B=N_E$ for the adaptive and fixed-rate schemes for $\mathcal{S}^{\text{th}} \in \{1,0.3,0.1\}$. We can see that, as the number of apertures increases for all nodes, the maximum obtained EST also increases. This can be explained  by the fact that the diversity order in the legitimate channel increases faster than that seen in the eavesdropper channel. Curiously, the EST is approximately the same for the adaptive scheme with $N_A=N_B=N_E=5$ and $\mathcal{S}^{\text{th}}=0.3$, and for the fixed-rate scheme with $N_A=N_B=N_E=10$ and $\mathcal{S}^{\text{th}}=1.0$. This implies that, in a scenario with five apertures per node, the adaptive scheme allows to restrict the SOP to be as low as $30\%$, while still achieving the same EST performance as the unconstrained fixed-rate scheme with ten apertures per node.
%for $\mathcal{S}^{\text{th}}$ and $N_A~=~N_B~=~N_E~=~1$, approximately same EST is obtained for both the adaptive scheme with $\mathcal{S}^{\text{th}}~=~0.01$ and for the fixed-rate with $\mathcal{S}^{\text{th}}~=~1.0$. This result indicates that, in a scenario with a single aperture per node, using the adaptive scheme allows the system to restrict the maximum secrecy outage probability of only $1\%$, while achieving the same EST as the fixed-rate in an unconstrained scenario.
%From Fig.~\ref{fig:Est_N_Adaptive_FixedRate}, it is also clear that as the amount of apertures increase for all nodes, besides the increase in the obtained EST, the system is able to operate in a much lower SOP without losing a significant performance when compared to the unconstrained scenario.
%that the diversity order in the legitimate channel (given by $N_A\cdot N_B$) increases faster than that seen in the eavesdropper channel (given by $N_E$).

\begin{figure}[!t]
\centering
\includegraphics[width=240px]{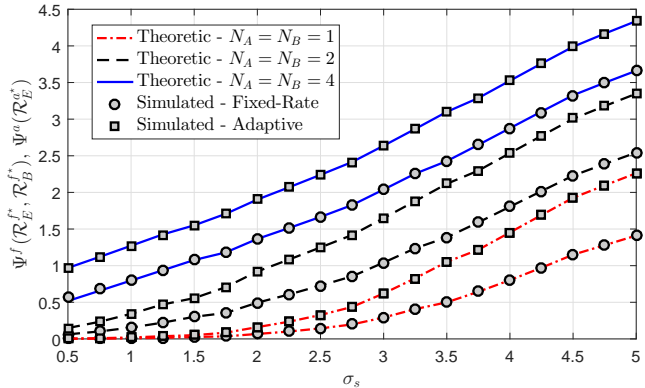}
\caption{$\Psi^f(\mathcal{R}^{f^\star}_{E},\mathcal{R}^{f^\star}_{B})$, $\Psi^a(\mathcal{R}^{a^\star}_{E})$ versus $\sigma_s$ for fixed-rate and adaptive transmission schemes with $N_E~=~2$ and $\mathcal{S}^{\text{th}} = 0.2$.}
\label{fig:Est_SigmaS_Adaptive_FixedRate}
\end{figure}

Finally, in Fig.~\ref{fig:Est_SigmaS_Adaptive_FixedRate} we present the EST versus $\sigma_s$ for the adaptive and fixed-rate schemes, with $\mathcal{S}^{\text{th}} = 0.2$. One can see that, as the standard deviation of pointing error displacement increases, the EST for both schemes increases. As seen in Fig.~\ref{fig:EST_RE_Adaptive}, this is due to the fact that the increase in $\sigma_s$ decreases the capacity of the eavesdropper channel. 
%Similar to that seen in Fig.~\ref{fig:Est_Sth_Adaptive_FixedRate}, one can also see that the adaptive scheme outperforms the fixed-rate scheme for all values of $\sigma_s$, which means that it is able to achieve a better EST independent of the distance of between Bob and Eve. \commbr{essa conclusão é meio que óbvia...}

\section{Final Comments} \label{sec:conclusions}

In this work, we characterized the MIMOME performance for coherent FSO transmissions. A threat scenario where Eve is near Bob was investigated, meaning that the difference of SNR seen at Bob and Eve is affected not only by $N_A$, $N_B$ and $N_E$, but also by the pointing errors due to misalignment between Alice and Eve. By adopting the EST with secrecy constraints as the performance metric, the optimal rates for the adaptive and fixed-rate schemes were obtained. Numerical results confirmed the accuracy of the mathematical derivations. Moreover, our analytical and simulation results demonstrated that, independent of the maximum allowed SOP, the EST for the adaptive scheme outperforms that obtained using the fixed-rate transmission scheme for coherent FSO communications. Finally, we also show that a significant gain is achieved when adding multiple apertures, and that the overall EST is also dependent on the distance between Eve and Bob. Future works include the analysis of the EST with secrecy constraints using non-coherent reception, in which the channel capacity changes significantly, and the use of relays.
\appendices

\section{Proof of Lemma~\ref{lemma1}} \label{appendix:lemma1}
In order to obtain~\eqref{eq:sop}, we first resort to the fact that the probability from~\eqref{eq:prsop} can be rewritten as
\begin{equation}\label{eq:provaprsop1}
\mathcal{S}(\mathcal{R}_E) = \Pr\left\{ Z \sum _{n=1}^{N_E}X_nY_n > \frac{N_0\left(2^{\mathcal{R}_E}-1\right)}{A_0} \right\},
\end{equation}where $X, Y \sim \Gamma \left( \cdot \right)$ are, respectively, the large-scale and small-scale parameters of the gamma-gamma random variable, both of which are gamma distributed with shape parameter $k$ inversely proportional to the scale parameter $\theta$, i.e, $\theta = \frac{1}{k}$, and $Z$ represents the random variable due to pointing error which the pdf is given by~\eqref{eq:perrorpdf}. Due to the spatial proximity of the apertures and the inherent LOS nature of FSO systems,  the large-scale can be assumed equal to all receive apertures~\cite{Balsells.14}, such that \eqref{eq:provaprsop1} can be rewritten as
\begin{equation}\label{eq:provaprsop3}
\mathcal{S}(\mathcal{R}_E) = \Pr\left\{ Z X \sum _{n=1}^{N_E}Y_n > \frac{N_0\left(2^{\mathcal{R}_E}-1\right)}{A_0} \right\}.
\end{equation}
Using the summation property, where the summation of a gamma variable with shape $k$ and scale $\theta$ can be expressed as a single gamma variable, where the shape parameter is the sum of all shape parameters~\cite{Moschopoulos.1985}, i.e., $\sum _{n=1}^{N_E} X_n \sim \Gamma \left( \sum _{n=1}^{N_E} k_n, \theta \right)$, \eqref{eq:provaprsop3} can be rewritten as
\begin{equation}\label{eq:provaprsop4}
\begin{split}
\mathcal{S}(\mathcal{R}_E) = \Pr \left\{ Z X_{\alpha } \frac{1}{N_E} Y_{\beta } > \frac{N_0\left(2^{\mathcal{R}_E}-1\right)}{N_EA_0} \right\},\\
\end{split}
\end{equation}where $X_{\alpha} \sim \Gamma \left( \alpha, \frac{1}{\alpha} \right)$ and $Y_{\beta} \sim \Gamma \left( N_E \beta, \frac{1}{\beta} \right)$. In order to obtain $\theta = \frac{1}{k}$ as proposed in~\cite{Habash2001} and used in~\eqref{eq:pdfgammagamma}, we resort to the scale property, where the product of a gamma variable by a constant can be rewritten as a gamma variable where the scale parameter $\theta$ is the product of the original scale by the constant, i.e., $cX \sim \Gamma(k, c\theta)$ and, thus%, \eqref{eq:provaprsop4} can be rewritten as
\begin{equation}\label{eq:provaprsop5}
\mathcal{S}(\mathcal{R}_E) = \Pr \left\{ Z X_{\alpha} Y_{\beta_{E}} > \frac{N_0\left(2^{\mathcal{R}_E}-1\right)}{N_EA_0} \right\},
\end{equation}where $Y_{\beta_{E}}~\sim~\Gamma \left( N_E \beta, \frac{1}{N_E \beta} \right)$.
The pdf of $Z X_{\alpha} Y_{\beta_{E}}$ is given by~\eqref{eq:pdfggp}, such that the correspondent CDF can be obtained as
\begin{equation}
\begin{split}
F_{k}^{\gamma\gamma p}(x) = \int_0^x f^{\gamma\gamma p}(I) dI,
\end{split}
\end{equation} which results in~\eqref{eq:CDFGammaGammaP} and can be used directly to obtain~\eqref{eq:sop}, concluding the proof.

%\begin{equation} \label{eq:sa}
%\mathcal{S}(\mathcal{R}_E)  = 1 - F_{E}^{\gamma\gamma}(\mathcal{X}_E).
%\end{equation}
%We can now use the CDF of a gamma-gamma random variable, which is given by~\eqref{eq:CDFGammaGamma}, in order to obtain~\eqref{eq:sa} and, replacing~\eqref{eq:sa} in \eqref{eq:effective_secrecy_throughput_a}, we obtain~\eqref{eq:sop} concluding the proof.
%Finally, in a similar approach as described in Lemma~\ref{lemma1}, we use~\eqref{eq:CDFGammaGamma} to obtain~\eqref{eq:lemma3}
% and is represented in the right-hand side of~\eqref{eq:sa}

\section{Proof of Theorem~\ref{theo1}} \label{appendix:theo1}
%, and $X^{\text{AP}}_k \sim \Gamma \left( k^{\text{AP}}_k, \theta^{\text{AP}}_k \right)$
In order to obtain~\eqref{eq:optimal_r_a}, we first resort to the fact that, according to~\cite{Ahmadi2009}, a gamma-gamma random variable can be approximated by a gamma variable $X^{\text{AP}}_k$ with shape $k^{\text{AP}}_k$ and scale $\theta^{\text{AP}}_k$ parameters given, respectively, by~\eqref{eq:kapk} and~\eqref{eq:thetaapk}. Thus,~\eqref{eq:provaprsop5} can be approximated as
\begin{equation}\label{eq:theo1prof1}
\begin{split}
\mathcal{S}(\mathcal{R}_E) &\approx \Pr \left\{ Z X^{\text{AP}}_E > \frac{N_0\left(2^{\mathcal{R}_E}-1\right)}{N_EA_0} \right\}\\
&\approx 1 - F_{k,AP}^{\gamma\gamma p}\left(\frac{N_0\left(2^{\mathcal{R}_E}-1\right)}{N_EA_0}\right),\\
%Q\left(k_E^{\text{AP}},0,\frac{\frac{N_0\left(2^{\mathcal{R}_E}-1\right)}{N_EA_0}}{\theta_E^{\text{AP}}}\right).
\end{split}
\end{equation} where $F_{k,AP}^{\gamma\gamma p}(\cdot)$ can be easily obtained as
\begin{equation}\label{eq:theo1prof2}
\begin{split}
&F_{k,AP}^{\gamma\gamma p}(x)~=\\
&\frac{\left(\frac{x}{\theta _E^{\text{AP}}}\right)^{k_E^{\text{AP}}} E_{-k_E^{\text{AP}}+\frac{\omega _e^2}{4 \sigma _s^2}+1}\left(\frac{x}{\theta _E^{\text{AP}}}\right)-\Gamma \left(k_E^{\text{AP}},\frac{x}{\theta _E^{\text{AP}}}\right)+\Gamma \left(k_E^{\text{AP}}\right)}{\Gamma \left(k_E^{\text{AP}}\right)}.\\
\end{split}
\end{equation}
By setting $\delta \Psi^{a}(\mathcal{R}_E)/ \delta \mathcal{R}_E = 0$ and solving to $\mathcal{R}_E$, we obtain the stationary point of $\Psi^{a}(\mathcal{R}_E)$, which is given by~\eqref{eq:theo1}. We have that, similarly to~\cite{yan.2015}, an analysis of identifying stationary points via~\eqref{eq:theo1} and $\delta^2 \Psi^{a}(\mathcal{R}_E)/ \delta \mathcal{R}^2_E$ is not tractable. Noting that the probability from~\eqref{eq:theo1prof1} is a monotonically decreasing function of $\mathcal{R}_E$ (for $\mathcal{R}_E<C_B$) and following a similar approach as presented in~\cite{yan.2015}, we instead investigate through simulations and numerical calculations the nature of the stationary points, finding that \eqref{eq:effective_secrecy_throughput_a} is concave or semi-concave with only one stationary point for all tested simulations. This is in agreement with the monotonically decreasing behavior of~\eqref{eq:theo1prof1}. Thus, we find that the stationary points given by~\eqref{eq:theo1} always identify the local maximum in the simulations, as presented in Section~\ref{sec:numerical_results}.

Resorting to the fact that $\mathcal{S}(\mathcal{R}_E)$ is a monotonically decreasing function of $\mathcal{R}_E$\footnote{ This can be easily proved by showing that $\partial \mathcal{S}(\mathcal{R})/\partial \mathcal{R}> 0$, $\forall \, \mathcal{R}<C_B$.},
one can see that the redundancy rate at a given threshold $\mathcal{S}^{\text{th}}$ is the minimum redundancy rate allowed. Using \eqref{eq:theo1prof1}, one can find the inverse function with respect to $\mathcal{S}^{\text{th}}$, which is given by \eqref{eq:optimal_rOeth_a}. Noting that $\Psi^a(\mathcal{R}_E)$ increases for $\mathcal{R}_E < \mathcal{R}^{a^\star}_{E,u}$ and decreases for $\mathcal{R}_E > \mathcal{R}^{a^\star}_{E,u}$, we have that, without constraints, $\mathcal{R}^{a^\star}_{E,u}$ represents the maximum redundancy rate, in the sense that any value different than $\mathcal{R}^{a^\star}_{E,u}$ will result in a lower value of $\Psi^{a}(\mathcal{R}_E)$.
Noting that $\mathcal{R}^{a^\star}_E$ cannot be smaller than~\eqref{eq:optimal_rOeth_a} (due to the SOP constraint), one can conclude that $\mathcal{R}^{a^\star}_{E,u}$ is the maximum value between \eqref{eq:theo1} and \eqref{eq:optimal_rOeth_a}, which is given by \eqref{eq:optimal_r_a}.
%
%\section{Proof of Lemma 2} \label{appendix:lemma3}
%
%In order to obtain~\eqref{eq:lemma2}, we first resort to the fact that the probability from~\eqref{eq:probTranFixed} can be rewritten as
%\begin{equation}\label{eq:provaprtx1fixed}
%\begin{split}
%&\text{Pr}^f_{\text{tx}}(\mathcal{R}_B) \\
%&= \Pr\left\{ \max_{\substack{1\leq i\leq n_A}} \left(\sum _{n=1}^{N_B}X_{i,n}Y_{i,n} \right) > \left( \frac{N_BN_0\left(2^{\mathcal{R}_B}-1\right)}{A_0^2}\right) \right\} \\
%&= \Pr\left\{ \sum _{n=1}^{N_B}X_nY_n  > \left( \frac{N_BN_0\left(2^{\mathcal{R}_B}-1\right)}{A_0^2}\right) \right\}^{N_A}. \\
%%\text{Pr}^f_{\text{tx}}(\mathcal{R}_B) = \Pr\left\{ \sum _{n=1}^{N_B}X_nY_n> \sqrt{\left( \frac{N_BN_0\left(2^{\mathcal{R}_B}-1\right)}{A_0^2}\right)} \right\},
%\end{split}
%\end{equation}
%
%Using a similar approach as presented in Appendix~\ref{appendix:lemma1}, the probability presented in~\eqref{eq:provaprtx1fixed} can be approximated as
%
%\begin{equation}\label{eq:provaprtx2fixed}
%\text{Pr}^f_{\text{tx}}(\mathcal{R}_B) \approx \Pr \left\{ X_{\alpha_B} Y_{\beta_B} > \sqrt{\frac{N_0\left(2^{\mathcal{R}_B}-1\right)}{N_B A_0^2} } \right\}^{N_A},
%\end{equation}
%where $X_{\alpha_B}~\sim~\Gamma \left( N_B \alpha, \frac{1}{N_B \alpha} \right)$ and $Y_{\beta_B}~\sim~\Gamma \left( N_B \beta, \frac{1}{N_B \beta} \right)$. In order to obtain~\eqref{eq:lemma2}, we can then use the CDF of a gamma-gamma random variable, which is given by~\eqref{eq:CDFGammaGamma}, concluding the proof.

\section{Proof of Lemma~\ref{lemma3}} \label{appendix:lemma3}

Using~\eqref{eq:effective_secrecy_throughput_f}, the values of $(\mathcal{R}_E,\mathcal{R}_B)$ that that jointly maximize the EST for the fixed-rate scheme can be written as
\begin{equation}
\begin{split}
(\mathcal{R}_{E}^{f^\star},\mathcal{R}_{B}^{f^\star}) = \argmax_{\substack{0 < \mathcal{R}_{B}\\ 0 < \mathcal{R}_{E} < \mathcal{R}_{B}}} \Psi^{f}(\mathcal{R}_E, \mathcal{R}_B).
\end{split}
\end{equation}

Replacing~\eqref{eq:sop} and~\eqref{eq:lemma2} in~\eqref{eq:effective_secrecy_throughput_f}, and using the approximation of a gamma-gamma variable  by a gamma variable described in Appendix~\ref{appendix:theo1},~\eqref{eq:effective_secrecy_throughput_f} can be approximated as
\begin{equation} \label{eq:effective_secrecy_throughput_f2_aprox}
\begin{split}
\Psi^{f}&(\mathcal{R}_E, \mathcal{R}_B) \approx\\
&~(\mathcal{R}_B - \mathcal{R}_E) \left[1 - Q\left(k_B^{\text{AP}},0,\frac{\mathcal{X}_B}{\theta_B^{\text{AP}}}\right)\right]\, F_{k,AP}^{\gamma\gamma p}\left(\mathcal{X}_E\right),\\
\end{split}
\end{equation}and, setting the first-order partial derivative of~\eqref{eq:effective_secrecy_throughput_f2_aprox} with respect of $\mathcal{R}_{B}$ to zero, we have that
\begin{equation}\label{eq:provaprtx3fixed}
\begin{split}
&0 = C_2 \left(1-C_1^{N_A}\right)-\\
&\frac{C_2 N_0 2^{\mathcal{R}_B} \log (2) N_A \mathcal{R}~e^{-\frac{\mathcal{X}_B}{\theta^{\text{AP}}_B}} \left(\frac{\mathcal{X}_B}{\theta^{\text{AP}}_B}\right)^{k^{\text{AP}}_B-1} C_1^{N_A-1}}{\theta^{\text{AP}}_B N_b A_0 \Gamma \left(k^{\text{AP}}_B\right)},\\
\end{split}
\end{equation}
where $C_2 = Q\left(k_E^{\text{AP}},0,\frac{\mathcal{X}_E}{\theta_E^{\text{AP}}}\right)$ and $\mathcal{R} = \mathcal{R}_B-\mathcal{R}_E$. Solving~\eqref{eq:provaprtx3fixed} for $\mathcal{R}_E$, we obtain~\eqref{eq:lemma3eq1}. Using a similar approach, by setting the first-order partial derivative of~\eqref{eq:effective_secrecy_throughput_f2_aprox} with respect of $\mathcal{R}_{E}$ to zero and solving to $\mathcal{R}_B$, we obtain~\eqref{eq:lemma3eq2}.
%\begin{equation}\label{eq:provaprtx4fixed}
%\begin{split}
%0 =& -(1- C_1^{N_A}) C_2 + \\
%& \frac{(1- C_1^{N_A}) N_0 2^{\mathcal{R}_E} \log (2) \mathcal{R}~e^{-\frac{\mathcal{X}_E}{\theta^{\text{AP}}_E}} \left(\frac{\mathcal{X}_E}{\theta^{\text{AP}}_E}\right)^{k^{\text{AP}}_E-1}}{N_E A_0 \theta^{\text{AP}}_E \Gamma \left(k^{\text{AP}}_E\right)}.\\
%\end{split}
%\end{equation}
%where $C_4 = 1- F_{E}^{\gamma\gamma}(\mathcal{X}_E)$. Solving~\eqref{eq:provaprtx4fixed} to $\mathcal{R}_B$, we obtain~\eqref{eq:lemma3eq2}.
%Solving~\eqref{eq:provaprtx4fixed} to $\mathcal{R}_B$, we obtain~\eqref{eq:lemma3eq2}.
Based on Young's theorem~\cite{Youngs.1984}, similarly to that used in~\cite{yan.2015}, the Hessian matrix of~\eqref{eq:effective_secrecy_throughput_f2_aprox} is symmetric, and can be expressed as
\begin{align}
  \Hessian &=
  \begin{bmatrix}
    \frac{\partial^{2} \Psi^{f}(\mathcal{R}_E, \mathcal{R}_B)}{\partial \mathcal{R}^2_B} & \frac{\partial^{2} \Psi^{f}(\mathcal{R}_E, \mathcal{R}_B)}{\partial \mathcal{R}_B \partial \mathcal{R}_E } \\
    \frac{\partial^{2} \Psi^{f}(\mathcal{R}_E, \mathcal{R}_B)}{\partial \mathcal{R}_E \partial \mathcal{R}_B } & \frac{\partial^{2} \Psi^{f}(\mathcal{R}_E, \mathcal{R}_B)}{\partial \mathcal{R}^2_E}
  \end{bmatrix} =
  \begin{bmatrix}
    \mathcal{A} & \mathcal{B} \\
    \mathcal{B} & \mathcal{C}
  \end{bmatrix}.
\end{align}

For $\mathcal{A} < 0$ and $\mathcal{A} \cdot \mathcal{C} - \mathcal{B}^2 > 0$, then $(\mathcal{R}_{E,u}^{f^\star},\mathcal{R}_{B,u}^{f^\star})$ can be used to obtain the local maximum of $\Psi^{f}(\mathcal{R}_E, \mathcal{R}_B)$.

\section{Proof of Theorem ~\ref{theo2}} \label{appendix:theo2}

First, one must note that $\mathcal{S}(\mathcal{R}_E)$ is a monotonically decreasing function of $\mathcal{R}_E$, which means that, if $\mathcal{R}_{E,u}^{f^\star}~<~\mathcal{R}^{{th}^\star}_{E}$, then using $\mathcal{R}_{E,u}^{f^\star}$ will result in a SOP greater than the maximum allowed $\mathcal{S}^{\text{th}}$. Following a similar approach as described in Appendix~\ref{appendix:theo1}, one can conclude that the optimal value of $\mathcal{R}_E$ in a constrained scenario is given by the maximum between $\mathcal{R}_{E,u}^{f^\star}$ and $\mathcal{R}^{{th}^\star}_{E}$, which results in~\eqref{eq:theo2eq1}.

Noting that $\mathcal{R}_{B,u}^{f^\star}$ does not represent the optimal value of $\mathcal{R}_{B}$ when $\mathcal{R}_{E,u}^{f^\star} < \mathcal{R}^{{th}^\star}_{E}$, one must find the optimal value of $\mathcal{R}_{B}$ for a fixed value of $\mathcal{R}_E = \mathcal{R}^{{th}^\star}_{E}$. Similar to that presented in~\cite{yan.2015} and described in Appendix~\ref{appendix:theo1}, we have that the value of $\mathcal{R}_B$ that achieves the stationary point of $\Psi^{f}(\mathcal{R}_E, \mathcal{R}_B)$ is the optimal $\mathcal{R}_B$ for a fixed value of $\mathcal{R}_E$, which is obtained by replacing~\eqref{eq:sop} and~\eqref{eq:lemma2} in~\eqref{eq:effective_secrecy_throughput_f},  equating the first derivative to zero and solving for $\mathcal{R}_B$. It follows that the first derivative is given by
\begin{equation}
\begin{split}\label{eq:theo2profeq1}
&0 = (1-\mathcal{S}^{\text{th}}) \left(1-C_1^{N_A}\right)- \\
&\frac{N_0 2^{\mathcal{R}_B} (1-\mathcal{S}^{\text{th}}) \log (2) N_A \mathcal{R}~C_1^{N_A-1}  \left(\frac{\mathcal{X}_B}{\theta_B^{\text{AP}}}\right)^{k_B^{\text{AP}}-1}}{\exp\left(-\frac{\mathcal{X}_B}{\theta_B^{\text{AP}}}\right) \theta_B^{\text{AP}} N_b A_0 \Gamma \left(k_B^{\text{AP}}\right) },
\end{split}
\end{equation}yielding~\eqref{eq:theo2eq3} and concluding the proof.%and, equating~\eqref{eq:theo2profeq1} to zero, we obtain~\eqref{eq:theo2eq3}, concluding the proof.

%
%\section*{Acknowledgment}
%\label{secAcknowledgment}
%
%{\it \textcolor{blue}{
%The authors would like to acknowledge and warmly thank the reviewers whose comments significantly improved the content and presentation of this paper.
%}}

\bibliographystyle{IEEEtran}
\bibliography{IEEEabrv,biblio}

\end{document}